\documentclass{jpsj-suppl}
\usepackage{txfonts} %Please comment out this line unless the txfonts package is availabe in your LaTeX system.
\usepackage{graphicx}
\usepackage{epstopdf}
\usepackage{epsfig}

\title{Interaction of antiprotons with nuclei}

\author{Jaroslava \textsc{Hrt\'{a}nkov\'{a}}$^{1}$ and Ji\v{r}\'{i} \textsc{Mare\v{s}}$^{1}$}

\inst{$^{1}$Nuclear Physics Institute, 250 68 \v{R}e\v{z}, Czech Republic}

\email{hrtankova@ujf.cas.cz}

\recdate{May 23, 2016}

\abst{We report on self-consistent calculations of $\bar{p}$ quasi-bound states in selected nuclei performed within the relativistic mean-field (RMF) model employing $\bar{p}$ coupling constants tuned to 
reproduce $\bar{p}$-atom data. We confirmed considerable polarization of the nuclear core induced by the  antiproton. The $\bar{p}$ annihilation was treated dynamically, taking into account the reduced phase space 
in the nuclear medium as well as the compressed nuclear density. The energy available for the annihilation products was evaluated self-consistently, considering additional energy shift due to particle momenta in the ${\bar p}$-nucleus system. Next, we constructed the $\bar{p}$-nucleus optical potential using scattering amplitudes related to the latest version of the Paris $\bar{N}N$ potential. We explored energy dependence of the potential and the implications for $\bar{p}$-nucleus quasi-bound states.}

\kword{antiproton-nucleus interaction, antiproton annihilation, RMF model, Paris $\bar{N}N$ potential}

\begin{document}
\maketitle

\section{Introduction}
The $\bar{p}$-nucleus interaction has attracted renewed interest at the prospect of future experiments with $\bar{p}$ beams at FAIR \cite{FAIR}. The possibility of formation of $\bar{p}$--nuclear bound states has been extensively studied within the 
RMF approach in recent years~\cite{Mishustin, LMSG, jarka}. The considerations about their existence are  supported by a strongly attractive potential that $\bar{p}$ feels in the nuclear medium. 
However, the $\bar{p}$--nucleus interaction is dominated by $\bar{p}$ annihilation which significantly 
influences propagation of the antiproton in nuclear matter. 
The data from experiments with $\bar{p}$ atoms \cite{mares} and $\bar{p}$ scattering off nuclei at low energies \cite{antiNNinteraction} could be well fitted by a strongly attractive and strongly absorptive $\bar{p}$-nuclear  
optical potential, imaginary part of which outweighs the real part. Still, the phase space for the annihilation products is 
significantly suppressed for the antiproton deeply bound in the nuclear medium, which could lead to relatively long living $\bar{p}$ inside the nucleus \cite{Mishustin}.  

In this contribution, we present our fully self-consistent calculations of $\bar{p}$--nuclear bound states.  First, we discuss the $\bar{p}$-nucleus interaction derived within the RMF model \cite{Walecka} using G-parity transformed proton-meson coupling constants, properly scaled 
to fit $\bar{p}$-atom data~\cite{mares}. We demonstrate for selected nuclei dynamical effects in the nuclear core caused by the antiproton and the role of proper treatment of the $\bar{p}$ annihilation in the nuclear medium. Next, we analyze energy and density dependence of the in-medium $\bar{N}N$ scattering amplitudes constructed from the Paris $\bar{N}N$ potential~\cite{Paris2} and apply them in calculations of $\bar{p}$-nuclear quasi-bound states.     
 
In Section 2, we briefly introduce the underlying model. Few selected representative results of our calculations are discussed in Section 3 and conclusions are summarized in Section~4.

\section{Model}

The $\bar{p}$ interaction with nuclei is studied within the RMF approach. The (anti)nucleons interact among themselves by the exchange of the scalar ($\sigma$) and vector ($\omega_{\mu}$, $\vec{\rho}_\mu$) meson fields, and the massless photon field $A_{\mu}$. The standard Lagrangian density $\mathcal{L}_N$ for nucleonic sector is extended by the Lagrangian density $\mathcal{L}_{\bar{p}}$ describing the antiproton 
interaction with the nuclear medium:
\begin{equation} \label{LagDens}
\mathcal L_{\bar{p}}= \bar{\psi}_{\bar{p}} [i\gamma^\mu \partial_\mu \!-\! m_{\bar{p}} \!-\! g_{\sigma{\bar{p}}}{\sigma} \!-\! g_{\omega{\bar{p}}} \gamma_\mu{\omega}^\mu \!-\! g_{\rho{\bar{p}}}\gamma_\mu \vec{\tau}\cdot{\vec{\rho}}^\mu \!-\! e\gamma_\mu \frac{1}{2}(1+\tau_3) {A}^\mu]\psi_{\bar{p}} \; ,
\end{equation}
where $m_{\bar{p}}$ denotes the mass of the antiproton; $m_\sigma$, $m_\omega$, $m_\rho$ are the masses of the considered meson fields; 
$g_{\sigma{\bar{p}}}$, $g_{\omega{\bar{p}}}$, $g_{\rho{\bar{p}}}$ and $e$ are the $\bar{p}$ couplings to corresponding fields.  

The equations of motion for the hadron fields are derived within the variational principle employing the mean-field and no-sea approximations. The Dirac equations for nucleons and antiproton read:  
\begin{equation} \label{Dirac antiproton}
[-i\vec{\alpha}\vec{\nabla} +\beta(m_j + S_j) + V_j]\psi_j^{\alpha}=\epsilon_j^{\alpha} \psi_j^{\alpha}, 
\quad j=N,\bar{p}~,
\end{equation}
where
\begin{equation}
S_j=g_{\sigma j}\sigma, \quad V_j=g_{\omega j} \omega_0 + g_{\rho j}\rho_0 \tau_3 + e_j \frac{1+\tau_3}{2}A_0
\end{equation}
are the scalar and vector potentials, and $\alpha$ denotes single particle states. The Klein--Gordon equations for the boson fields acquire additional source terms due to the presence of $\bar{p}$:
\begin{equation}
\begin{split} \label{meson eq}
(-\triangle + m_\sigma^2+ g_2\sigma + g_3\sigma^2)\sigma&=- g_{\sigma N} \rho_{SN}-g_{\sigma \bar{p}} 
\rho_{S \bar{p}}~, \\
(-\triangle + m_\omega^2 +d\omega^2_0)\omega_0&= g_{\omega N}\rho_{VN} +g_{\omega \bar{p}} \rho_{V\bar{p}}~, \\
(-\triangle + m_\rho^2)\rho_0&= g_{\rho N}\rho_{IN} +g_{\rho \bar{p}}\rho_{I \bar{p}}~, \\
-\triangle A_0&= e_N \rho_{QN}+e_{\bar{p}}\rho_{Q\bar{p}}~,
\end{split}
\end{equation}
where $\rho_{\text{S}j}, \rho_{\text{V}j}, \rho_{\text{I}j}$ and $\rho_{\text{Q}j}$  are the scalar, 
vector, isovector, and charge densities, respectively. The values of the nucleon--meson coupling constants and meson masses are adopted from a particular RMF parametrization. In this work, we present results for the nonlinear RMF model TM1(2) \cite{Toki} for heavy (light) nuclei, the nonlinear NL-SH model \cite{nlsh} and the density dependent model TW99 \cite{TypelWolter}. In the case of the density dependent model, the couplings are a function of baryon density
\begin{equation}
 g_{iN}{\scriptstyle(\rho_{\text{V}N})}=g_{iN}{\scriptstyle(\rho_{\text{sat}})}f_i{\scriptstyle(x)}~, \quad i=\sigma, \omega, \rho~,
\end{equation}
where $x=\rho_{\text{V}N}\slash\rho_{\text{sat}}$. The system of the coupled Dirac  \eqref{Dirac antiproton}  and Klein--Gordon \eqref{meson eq} equations is solved fully self-consistently by iterative procedure.

\subsection{$\bar{p}$-nucleus interaction}

First, the $\bar{p}$--nucleus interaction is constructed within the RMF model from the $p$--nucleus interaction using the 
G-parity transformation: the vector potential generated by the $\omega$ meson changes its 
sign and becomes attractive. As a consequence, the total $\bar{p}$ potential would be excessively  
attractive. The G-parity transformation is surely a valid concept for long- and medium-range $\bar{p}$ potential, however, at short distances the $\bar{p}N$ interaction is dominated by strong annihilation.  
To take into account possible deviations from G parity due to the absorption as well as 
various many-body effects in the nuclear medium~\cite{Mishustin}, we introduce a scaling factor $\xi \in \langle0,1\rangle$ for the $\bar{p}$--meson 
coupling constants:
\begin{equation} \label{reduced couplings}
g_{\sigma \bar{p}}=\xi\, g_{\sigma N}, \quad g_{\omega \bar{p}}=-\xi\, g_{\omega N}, \quad g_{\rho 
\bar{p}}=\xi\, g_{\rho N}~.
\end{equation}

The $\bar{p}$ annihilation in the nuclear medium is described by the imaginary part of the optical potential in a `$t\rho$' 
form adopted from optical model phenomenology \cite{mares}:
\begin{equation}
2\mu {\rm Im}V_{\text{opt}}(r)=-4 \pi \left(1+ \frac{\mu}{m_N}\frac{A-1}{A} 
\right){\rm Im}b_0 \rho(r)~,
\end{equation}
where $\mu$ is the $\bar{p}$--nucleus reduced mass. The density $\rho(r)$ is evaluated dynamically 
within the RMF model, while the parameter Im$b_0=1.9$~fm as well as $\xi=0.2$ are determined by 
fitting the $\bar{p}$ atom data \cite{mares}. It is to be noted that the effective scattering length Im$b_0$ describes the $\bar{p}$ absorption at threshold. In the nuclear medium, the energy available for annihilation is reduced due to the binding of the antiproton and nucleon. Consequently, the phase 
space accessible to  annihilation products is suppressed. The absorptive $\bar{p}$ potential then acquires the form
\begin{equation} \label{im_pot}
 {\rm Im}V_{\bar{p}} (r,\sqrt{s},\rho)=\sum_{\text{channel}} B_c f_{\text{s}}(\sqrt{s}) {\rm Im}V_{\text{opt}}(r)~,
\end{equation}
where $f_s(\sqrt{s})$ is the phase space suppression factor and $B_c$ is the branching ration for a given channel (see ref. \cite{jarka} for details).
\bigskip

Next, the  S-wave $\bar{p}N$ scattering amplitudes derived from the latest version of the 
Paris $\bar{N}N$ potential~\cite{Paris2} are used to develop a $\bar{p}$ optical potential. 
The amplitudes are modified using the multiple scattering approach of Wass et al. \cite{wrw} in order to account for Pauli correlations in the medium. The in-medium isospin 1 and 0 amplitudes are of the form
\begin{equation}
F_{1}=\frac{f_{\bar{p}n}(\delta \sqrt{s})}{1+\frac{1}{4}\xi_k \frac{\sqrt{s}}{m_N} f_{\bar{p}n}(\delta \sqrt{s}) \rho}~, \qquad F_{0}=\frac{[2f_{\bar{p}p}(\delta \sqrt{s})-f_{\bar{p}n}(\delta \sqrt{s})]}{1+\frac{1}{4}\xi_k \frac{\sqrt{s}}{m_N}[2f_{\bar{p}p}(\delta \sqrt{s}) - f_{\bar{p}n}(\delta \sqrt{s})] \rho}~. 
\end{equation} 
Here, $f$ denotes the free-space amplitude as a function of $\delta \sqrt{s}=\sqrt{s}-E_{\text{th}}$; 
$\rho$ is the nuclear core density distribution and $\xi_k$ is defined as follows
%\begin{equation}
%\xi_k=\frac{9\pi}{p_f^2}\frac{1}{24}\left(4(q^2+6)+q(-2q(q^2-12)\ln(q)+(q+4)(q-2)^2\ln(q-2)+(q-4)(q+2)^2\ln(q+2))\right)~.
%\end{equation}
\begin{equation}
 \xi_k=\frac{9\pi}{p_f^2}\,4\! \int_0^{\infty} \frac{dt}{t} \exp(iqt)j_1^2(t)~,
\end{equation}
where $j_1(t)$ is Spherical Bessel function, $p_f$ is Fermi momentum, $q=k/p_f$ and $k=\sqrt{E_{\bar{p}}^2-m_{\bar{p}}^2}$ is the $\bar{p}$ momentum. The S-wave optical potential is expressed as
\begin{equation} \label{parisS}
	2E_{\bar{p}}V_{\text{opt}}=-4\pi \frac{\sqrt{s}}{m_N}\left(F_0\frac{1}{2}\rho_p + F_1\left(\frac{1}{2}\rho_p+\rho_n\right)\right)~,
\end{equation}
  where $\rho_p$ ($\rho_n$) is the proton (neutron) density distribution and the factor 
$\sqrt{s}/m_N$ transforms the in-medium amplitudes to the $\bar{p}$-nucleus frame.
\bigskip

The energy available for the $\bar{p}$ annihilation in the nuclear medium is given by Mandelstam variable
\begin{equation} \label{s}
s=(E_N + E_{\bar{p}})^2 - (\vec{p}_N + \vec{p}_{\bar{p}})^2~,
\end{equation} 
where $E_N=m_N - B_{Nav}$ and $E_{\bar{p}}=m_{\bar{p}}-B_{\bar{p}}$, with $B_{Nav}$ being the average 
binding energy per nucleon and $B_{\bar{p}}$ the $\bar{p}$ binding energy.  
In the two-body c.m. frame $\vec{p}_N + \vec{p}_{\bar{p}} = 0$ and Eq.~\eqref{s} reduces to
\begin{equation} \label{Eq.:M}
\sqrt{s}=~m_{\bar{p}}+m_{N}-B_{\bar{p}}-B_{Nav}~~~(\text{M}).
\end{equation}
However, when the annihilation of the antiproton with a nucleon takes place in a nucleus, the momentum dependent term in Eq.~\eqref{s}
is no longer negligible~\cite{s} and provides additional downward energy shift. Taking into account averaging over the angles $(\vec{p}_N + \vec{p}_{\bar{p}})^2 \approx \vec{p}_N^{~2}+\vec{p}_{\bar{p}}^{~2}$,
Eq.~\eqref{s} can be rewritten as \begin{equation} \label{Eq.:J}
 \sqrt{s}= E_{th} \left(\!1-\frac{2(B_{\bar{p}} + B_{Nav})}{E_{th}} + \frac{(B_{\bar{p}}+ B_{Nav})^2}{E_{th}^2} - \frac{1}{E_{th}}T_{\bar{p}} - \frac{1}{E_{th}}T_{Nav} \!\right)^{1/2}~~~(\text{J}),
\end{equation}
where $T_{Nav}$ is the average kinetic energy per nucleon 
and $T_{\bar{p}}$ represents the $\bar{p}$ kinetic energy. The kinetic energies were calculated as
the expectation values of the kinetic energy operator ${T}_j=-\frac{\hbar^2}{2 m_j^{(*)}} \triangle$, where $m^*_j=m_j-S_j$ is the
(anti)nucleon reduced mass.

For comparison, we applied in our calculations also another form of $\sqrt{s}$ which was originally 
used in the studies of $K^-$-nuclear potentials \cite{s, kaony}. The momentum dependence in $\sqrt{s}$ was transformed into the density dependence. The nucleon kinetic energy was approximated within the Fermi gas model by $T_N(\frac{\rho}{\rho_0})^{2/3}$, where $T_N=23$~MeV,
and the kaon kinetic energy was expressed within the local density approximation by $T_K \approx -B_K-{\rm Re}{\cal V}_K(r)$, where ${\cal V}_K = V_K + V_{\rm C}$ and $V_{\rm C}$ is the $K^-$ Coulomb potential, which led
to the expression
\begin{equation} \label{Eq.:K}
 \sqrt{s}= m_N + m_K - B_{Nav}- \xi_N B_K + \xi_K {\rm Re}{\cal V}_K(r) - \xi_N T_N(\frac{\rho}{\rho_0})^{2/3}~~~(\text{K}),
\end{equation}
where $\xi_{N(K)}={m_{N(K)}}/(m_N+m_K)$.

\section{Results}
We applied the RMF formalism introduced above in calculations of $\bar{p}$ quasi-bound states in various nuclei. First, we did not consider the $\bar{p}$ absorption and studied dynamical effects in a nucleus due to the presence of $\bar{p}$. 
We confirmed a large polarization of the nuclear core caused by the antiproton. The $\bar{p}$ energies calculated dynamically are substantially larger than those calculated statically (i.e., 
without $\bar{p}$ source terms in the r.h.s of Klein-Gordon equations \eqref{meson eq}). Moreover, the polarization of the nuclear core varies with the applied RMF model due to different values of nuclear compressibility \cite{jarka}.

\begin{figure}[b]
 	\begin{center}
		\includegraphics[width=0.4\textwidth]{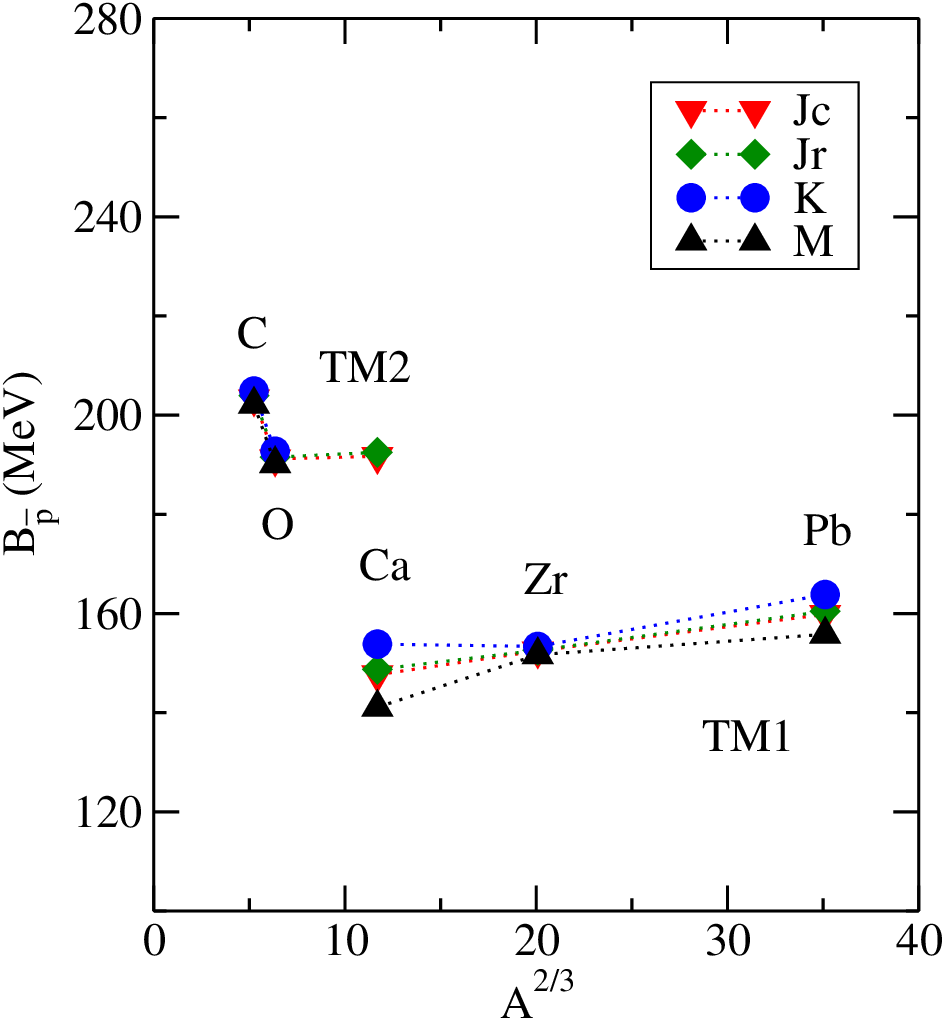} \hspace{8pt}
 		\includegraphics[width=0.4\textwidth]{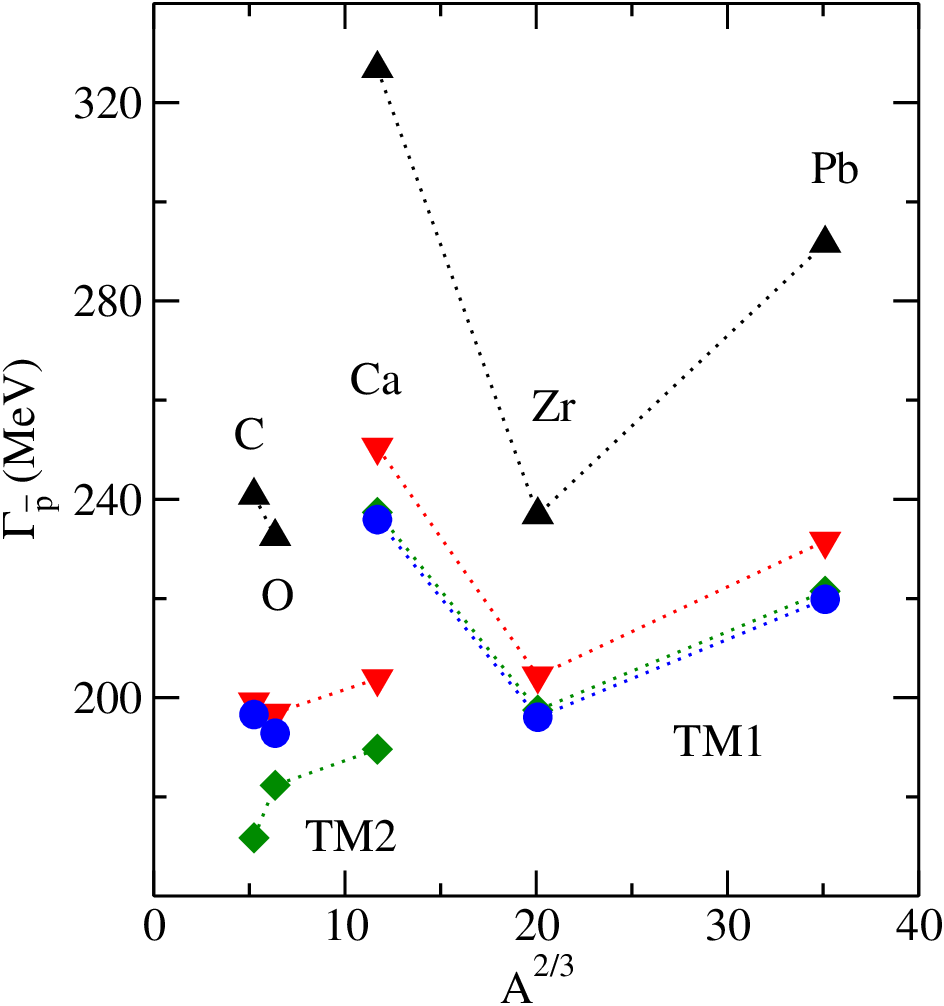}
\end{center}
 		\caption{\label{fig.:energie} Binding energies (left panel) and widths (right panel) of $1s$ $\bar{p}$-nuclear states in selected nuclei, calculated dynamically using the TM models and different forms of $\sqrt{s}$ (see text for details).}
 \end{figure}
 
The $\bar{p}$ absorption in a nucleus was described by the imaginary part of phenomenological optical potential (Eq.~\eqref{im_pot}) and treated self-consistently. In Fig. \ref{fig.:energie}, we present $1s$ $\bar{p}$ binding energies (left panel) and widths (right panel) in selected nuclei calculated in the TM models for different forms of $\sqrt{s}$. The $\bar{p}$ energies do not deviate much from each other. However, the $\bar{p}$ widths, which are sizable in all nuclei considered, exhibit much larger dependence on the applied form of $\sqrt{s}$. The largest widths are predicted for $\sqrt{s}=$~M in the two body frame 
(see Eq.~(13)). The ${\bar p}$ widths are significantly reduced after including the momentum dependent terms in $\sqrt{s}$ (Eq.~(14)).
\begin{figure}[t]
 	\begin{center}
 		\includegraphics[width=0.4\textwidth]{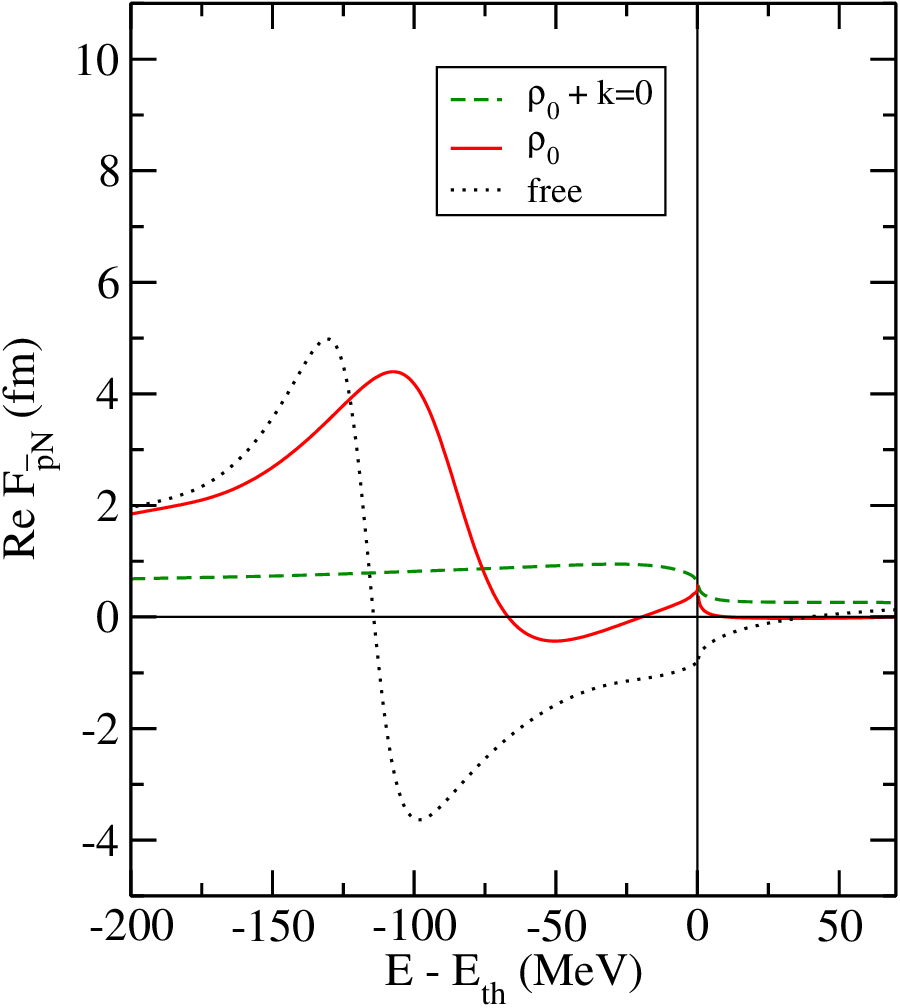} \hspace{8pt}
 		\includegraphics[width=0.4\textwidth]{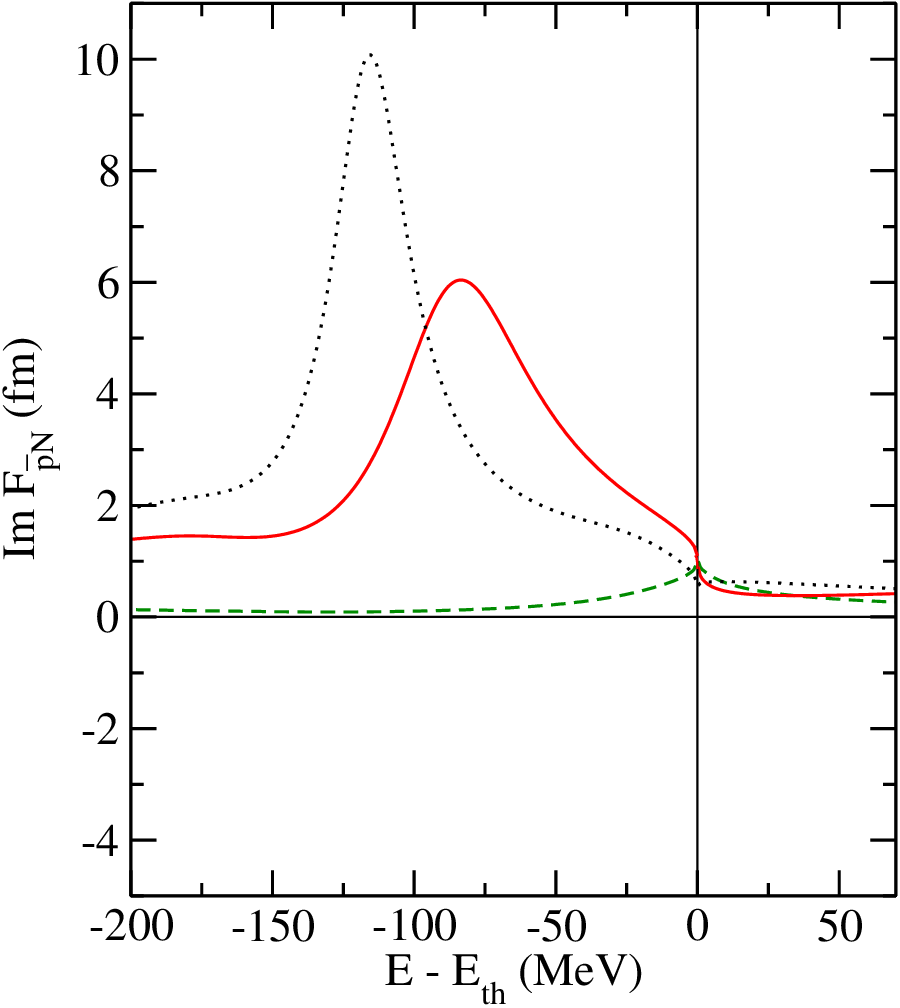}
\end{center}
 		\caption{\label{fig.:ampl}Energy dependence of the Paris 09 $\bar{p}N$ S-wave amplitudes: in-medium (Pauli blocked) amplitudes for $\rho_0=0.17$~fm$^{-3}$ (solid line) and for $\rho_0=0.17$~fm$^{-3}$ but $\bar{p}$ momentum $k=0$ ($\xi_k =1$ in Eq.(10)) (dashed line) are compared with the free-space amplitude (dotted line).}
 \end{figure} 
In order to study the effects of the medium, the (anti)nucleon kinetic energies were calculated with constant as well as reduced (anti)nucleon masses. As a result, the kinetic energies calculated with reduced masses ($\sqrt{s}=$~Jr) are larger and consequently the $\bar{p}$ widths are smaller than those calculated using constant masses ($\sqrt{s}=$~Jc). The $\bar{p}$ widths calculated using $\sqrt{s}=$~K (Eq.~(15)) and Jr are comparable.

Next, we calculated the $\bar{p}$-nuclear quasi-bound states using the optical potential derived from 
the  S-wave scattering amplitudes of the Paris $\bar{N}N$ potential (Eq.~\eqref{parisS}). In Fig.~\ref{fig.:ampl}, the energy dependence for the free-space and in-medium $\bar{p}N$ amplitudes for $\rho_0=0.17$~fm$^{-3}$ is shown. The peaks of the in-medium amplitudes (solid line) are lower in comparison with the free amplitudes and shifted towards threshold. In the case of the in-medium amplitude evaluated for the $\bar{p}$ momentum $k=0$ (dashed line), both the real and imaginary part are substantially reduced and 
become smooth in the whole energy region. 

\begin{figure}[h]
	\begin{center}
	\includegraphics[width=0.75\textwidth]{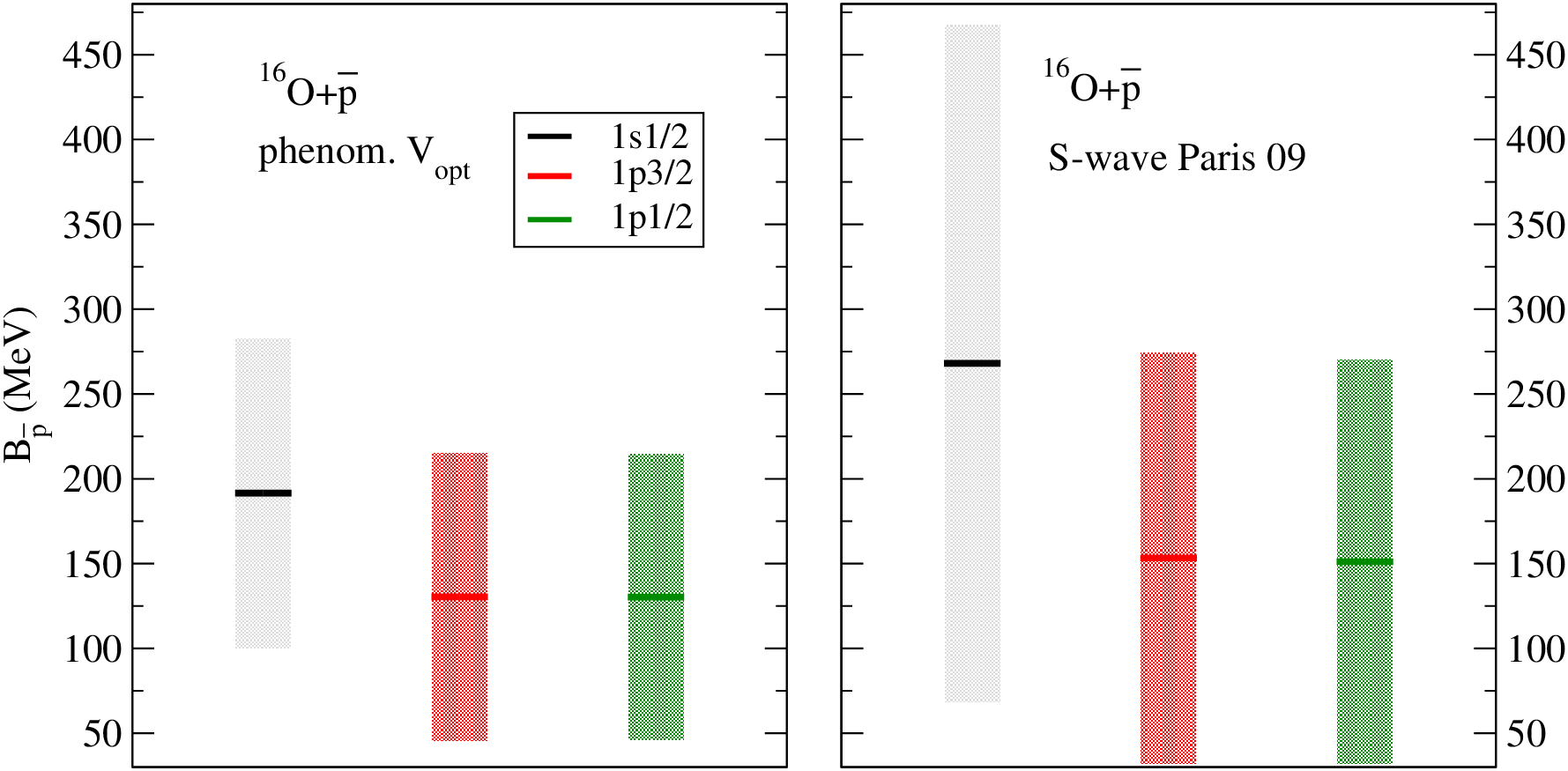}
\end{center}
	\caption{\label{fig.:kyslik} $1s$ and $1p$ binding energies (lines) and widths (boxes) of $\bar{p}$ in $^{16}$O calculated dynamically within the TM2 model for $\sqrt{s}=$ Jr with phenomenological $\bar{p}$ optical potential (left) and S-wave Paris potential (right).}
\end{figure}

Fig.~\ref{fig.:kyslik} shows the $1s$ and $1p$ binding energies and widths of the antiproton in $^{16}$O calculated dynamically  with 
the  S-wave Paris potential for $\sqrt{s}=$~Jr, compared with the phenomenological RMF approach. The S-wave Paris potential yields similar spectrum of $\bar{p}$ bound states as the phenomenological potential, however the $\bar{p}$ energies and widths are larger than those calculated within the RMF model, particularly in  
the $1s$ state. It is to be noted that the Paris $\bar{N}N$ potential contains a sizable $P$-wave interaction which should be taken into account. 
Calculations involving the P-wave term in the optical potential are currently in progress and will be published elsewhere.

\section{Conclusions}

In this work, $\bar{p}$-nucleus quasi-bound states in selected nuclei were studied. 
The $\bar{p}$-nucleus interaction was constructed using two different approaches: 
a) the RMF model with G-parity motivated $\bar{p}$ coupling constants, properly scaled to fit $\bar{p}$-atom data, and a 
phenomenological absorptive part;   
b) the model based on in-medium scattering amplitudes derived from the latest version of the Paris $\bar{N}N$ potential. 
We explored dynamical effects caused by the presence of the strongly interacting ${\bar p}$ in selected nuclei across the periodic table and confirmed sizable changes in the nuclear structure. The dependence of the $\bar{p}$ energies and widths on the applied form of 
$\sqrt{s}$ was discussed. We evaluated self-consistently additional downward energy shift due to the $\bar{p}$ and $N$ momenta, which 
leads to significant suppression of the $\bar{p}$ widths in the nuclear medium. However, the widths remain still sizable. 
We calculated the spectrum of $\bar{p}$ bound states in $^{16}$O using the S-wave Paris potential for the first time. 
The calculated $\bar{p}$ binding energies and widths are larger than those obtained by the RMF approach. 

\section*{Acknowledgements}
We wish to thank E. Friedman, A. Gal and S. Wycech for valuable discussions, and B. Loiseau for providing us with the $\bar{N}N$ amplitudes. This work was supported by the GACR Grant No. P203/15/04301S.

\end{document}